\documentclass[conference]{IEEEtran}
\usepackage{cite}
\usepackage{amsmath}
\usepackage{graphicx}
\usepackage{url}
\usepackage{algorithm}
\usepackage{algpseudocode}
\usepackage{multirow}

\usepackage{textcomp}
\usepackage{bm}
\usepackage{comment}
\usepackage{boldline}
\ifCLASSINFOpdf
\else
\fi
\begin{document}
\title{Tweets Can Tell: Activity Recognition using Hybrid Long Short-Term Memory Model}

\author
{
\IEEEauthorblockN{Renhao Cui, Gagan Agrawal, Rajiv Ramnath}
\IEEEauthorblockA{Department of Computer Science and Engineering\\
The Ohio State University\\
Columbus, Ohio\\
cui.182@osu.edu, agrawal.28@osu.edu, ramnath.6@osu.edu}
}

\maketitle

\baselineskip=0.989\normalbaselineskip  

\begin{abstract}

This paper presents techniques to detect the “offline” activity a person is engaged in when she is tweeting (such as dining, shopping or entertainment), 
in order to create a dynamic profile of the user, for uses such as  better targeting of advertisements. 
To this end, we propose a hybrid LSTM model for rich contextual learning, 
along with studies on the effects of applying and combining multiple LSTM based methods with different contextual features. 
The hybrid model is shown to outperform a set of baselines and state-of-the-art methods. 
Finally, this paper presents an orthogonal validation with a real-case application.
Our model generates an offline activity analysis for the followers of several well-known accounts, which is quite representative of the expected characteristics of these accounts.

\end{abstract}
\label{Activity Recognition}
\section{Introduction}

Precise real-time targeting of advertisements is essential to the success of online advertising. 
Social media platforms are able to build rich  profiles from the online presence of users by tracking activities such as participation, 
messaging and website visits. 
The important question we seek to address in this paper is,  {\em ``Can we tell what the user is actually doing when she tweets?''}
For example, is she dining, watching a movie, or studying in a library? 
By knowing the activities of a user, such as whether they visit restaurants or travel frequently,  more precisely targeted advertisements and marketing strategy can be directed to them. 

Social media posts of users are primarily driven by their interests.
Extracting these interests from  posts has been quite successful \cite{michelson2010discovering,kapanipathi2014user}. We now seek to unearth the offline activities that the user is engaged in when she posts, because these can provide a close to real-time view into the user.
As an example, building interest profiles may tell us that a user likes watching movies, so ads related to certain types of movies may evoke her attention. However, being able to detect  offline activities can tell us that a user is watching a movie right now, so ads related to popcorn and beer may be immediate appeal.
In other words, 
knowing the activity a user is engaged in can enable very effectively 
targeted advertising. 

\begin{table}[htb]
\caption{Sample Tweets with Reported Locations}
\footnotesize
\centering
\begin{tabular}{c|l|c|c}
\hline
&\textbf{Content}&\textbf{Location}&\textbf{Activity}\\
\hline\hline
1&Just Landed in Looondon&Airport&Traveling\\
\hline
\multirow{2}{*}{2}&We've been trapped in London&\multirow{2}{*}{Airport}&\multirow{2}{*}{Traveling}\\
&for 12 hours&&\\
\hline
\multirow{2}{*}{3}&Ready @Tomlovestorun1? I'm&\multirow{2}{*}{Airport}&\multirow{2}{*}{Traveling}\\
&not so sure&&\\
\hline
4&Happy national tequila day!&Night Club&Entertaining\\
\hline
\end{tabular}
\end{table}

Detecting a user's activity from a tweet could be difficult. To illustrate this, Table 1 shows a set of sample tweets along with their reported locations and their assigned activity labels.
The keyword ``landed'' in Tweet 1 is sufficient to identify the correct location of the user (airport) and her activity (traveling). 
Tweet 2 needs some inference to understand the situation of its author -- being stuck in a major transportation center. This situation can still be extracted from the content of the tweet.
Tweet 3 contains no information at all of its activity -- travelling.
Further, a naive model may identify the activity of Tweet 4 as dining, because the tweet talks about  a drink; however, the author is actually entertaining at a nightclub.
In fact, we have observed that it is quite common to post tweets with content that may clearly indicate one type of activity, while the author is actually engaged in a different type of activity. 

In other words, these examples show that the semantic content of a social media post does not, by itself, always provide meaningful information related to the activity that the author is engaged in while posting.
Additionally, user-reported locations are very useful in determining such activities. For example, \cite{yang2015modeling,liao2018predicting,lian2011collaborative} have shown correlation between activities and the check-in locations of the posts.
However, very few tweets contain such location information.

Our goal is, therefore, to build a model that is able to recognize user activities not only for the cases where a clear indicator exists in the content, 
but also for the ones where the activity information is latent and not directly usable.
Therefore, the model should work without the help of author-provided location information.

Continuing the motivation for this work, it should be clear that for tweets 3 and 4, their content alone is not sufficient to extract the correct offline activity, and additional context knowledge is needed.
For example, the additional knowledge of post time of tweet 4 (midnight) dramatically increases the possibility that the author is being entertained at a night club rather than eating at a restaurant.
The historical information is another contextual information. Thus, knowing that a post prior to tweet 3 is about heading home allows us to infer that the author sent this post while traveling.
Thus, we posit that in order to recognize offline activity, a richer contextual model is required, consisting of additional background information.

To show that such inference can be handled effectively, this paper focuses on the following research questions:
\begin{itemize}
\item How can we identify and appropriately label the offline activities of tweets?
\item What contextual information (i.e. other than the content) assists in recognizing activities?
\item How can we effectively recognize user activities using the contextual features?
\end{itemize}

We address these questions through novel techniques as well as enhancements to existing techniques.  
We start  by  using a Long Short-Term Memory (LSTM) network \cite{hochreiter1997long} to model only the content of tweets.
LSTM is designed to handling sequential data, and it has been shown to provide a reasonable performance on tweet classifications \cite{huang2016modeling,li2016tweet,wang2015predicting}.  
To further improve the model, we explore and analyze the inclusion of other contextual features with different variations of LSTM model.
Based on the analysis and comparison, we propose a hybrid LSTM model that properly handles the contextual features to improve the outcome. 
For evaluation, we create a labeled dataset by collecting tweets where users have reported their location.
For the activity classification task, our proposed model is able to reduce the error by 12\% over the content-only models and 8\% over the existing contextual models.

Finally, this paper presents an orthogonal validation towards the proposed hybrid model with a real-case application. 
Our model forms an analysis towards the activities of the followers of several well-known Twitter accounts, 
and the analysis demonstrates strong relationships to the expected characteristics of these accounts.
To the best of our knowledge, this is the first work that seeks to recognize offline activities using a author-independent model.
It is also the first work that looks into and compares different LSTM based models with respect to their abilities to work with contextual features. 

\section{Related Work}

User profiling on social media has been a popular area, and it is useful for  personalization, recommendation, and advertising.
Research has been conducted on user profiling based on the posts and interactions between the  users.
Rao {\em et al.} \cite{rao2010classifying} use linguistic features to profile users to extract gender, age, regional origin, and political orientation.
Lee {\em et al.} \cite{lee2014user} build a user profile model based on certain types of words to improve new recommendations.
Certain efforts~\cite{benevenuto2009characterizing,atig2014activity,malmgren2009characterizing} characterize users based on their online communication and web-page  visiting activities.
Detecting life events \cite{yen2018detecting,dickinson2016identifying} from tweets has also been addressed.  

The problem of  inference and prediction of real-life activities of users has not received much attention.  
So far, there are mainly two types of works on the extraction of offline activities of users: 
 recognition of the current activity (activity recognition) and prediction of a future activity (activity prediction).
Activity prediction  considers all features as 
historical data, while activity recognition 
focuses on 
current activities.
Early works on activity prediction \cite{ye2013s,noulas2011empirical,lian2011collaborative} relies on the history of check-in locations provided by the user.
Later, \cite{yang2015modeling} and \cite{liao2018predicting} add temporal information to the analysis of activities given location data.
None of the work utilize the post content of the users, which is the  major focus of our models.
Weerkamp {\em et al.} \cite{weerkamp2012activity} predict future activities by summarizing tweet topics where a future time frame is mentioned.
To recognize the current activities, Song {\em et al.} \cite{song2013collaborative} build a framework that incorporates the similarity measurement between the bag-of-word based classifiers of different users by comparing the decisions of the classifiers. 
It assumes that friends on social platforms are related in their activities.
Relation in user interest is quite common among friends, however, offline activities do not necessarily hold the same assumption.
In contrast, our belief is that contextual information provided by the same author is more relevant in recognizing offline activities.

For the task of text mining, LSTM \cite{hochreiter1997long} has been widely used for modeling sequential data.
Greff {\em et al.} \cite{greff2017lstm} perform a comparison across eight content-based LSTM variants, and demonstrate that these variants only have limited improvements.
To improve the performance, Bi-directional LSTM (BiLSTM)\cite{schuster1997bidirectional}  and LSTM with a Convolutional Neural Network (CNNLSTM)\cite{zhou2015c} are introduced to capture more appropriate information.
Recently, attention mechanisms are added to LSTM \cite{wang2016attention,liu2016learning} to strength the ability of handling long-dependencies.
In order to incorporate external information, Ghosh {\em et al.} \cite{ghosh2016contextual} build a contextual LSTM model that adds the contextual feature into the calculation of each gate function.
Yen {\em et al.} \cite{yen2018detecting} utilize a multi-task LSTM and include contextual information by simply concatenating the features.
Finally, hierarchical LSTM models are built \cite{zhou2016hierarchical,huang2016modeling} that stack LSTM models with different levels of sequential data.
In general, the effectiveness of each model is highly reliant on the input data and features; thus, none of the models appear good enough to work with all types of contextual data.
We look into the capabilities of several contextual models with respect to different contextual features, and come up with a hybrid model that takes advantage of the success of these models.
\section{Working with Contextual Features using LSTM}
In this section, we first describe the process of creating and assigning activity labels to tweets.
Then we show the work on exploring several models that are built based on LSTM to include contextual features.

\subsection{Activity Labeling}  


Similar to the labeling approaches of \cite{lian2011collaborative} and \cite{song2013collaborative}, 
we design an automatic labeling process that uses the reported location of the tweets to assign labels.
The reported location is highly predictive to the activities of the tweet.
Essentially, we categorize locations and use predefined rules to map locations to activities.
Note that we also create additional mapping rules to overcome errors brought by locations that could be involved in multiple activities.

\subsection{Contextual Learning with LSTM}
A typical approach to improve model performance is to include additional, and hopefully, more useful features.
We therefore look into several popular LSTM-based models that used contextual features including static features such as time of post, sequential features such as part-of-speech (POS) tags, and historical features such as the most recent tweets from the same author.
The sequence of POS tags helps to better understand the content, beginning with the positioning of words; while the timing of the post and historical tweets may be able to provide useful background knowledge of the target tweet.
Because the goal of the system is to provide real-time recognition of activities associated with a given target tweet, we only 
utilize tweets posted prior to the target tweet.
We do not include the topics of the tweet since they have been shown to be ineffective in past studies \cite{yang2014large}\cite{mehrotra2013improving}.


\subsubsection{Original LSTM}
Sequential models such as LSTM and Gated Recurrent Unit (GRU) \cite{cho2014learning} are ideal for text processing, because they consider the order and dependencies of tokens.
Since LSTM and GRU have comparable performance, we use LSTM as the baseline to improve by including contextual features.

A simplified architecture of the LSTM model used for a text classification problem is shown in Figure 1.
The output of the embedding layers is a sequence of vectors that represent the input sequence.
LSTM outputs a flat vector representation for the entire input sequence, and it is fed into another layer to generate the classification output.
For our activity recognition task, the tweet content is the input and  the activity label is the output.

\begin{figure}[htb]
\centering
\includegraphics[scale=0.6]{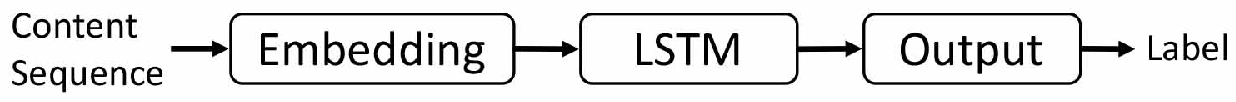}
\caption{LSTM for Text Classification}
\end{figure}

\subsubsection{Joint-LSTM} 

Similar to the idea of Yen {\em et al.} \cite{yen2018detecting}, 
we design Joint-LSTM (J-LSTM) model to concatenate the flat representations of the sequential input of content and contextual features before feeding it to the output layer.

\begin{figure}[htb]
\centering
\includegraphics[scale=0.4]{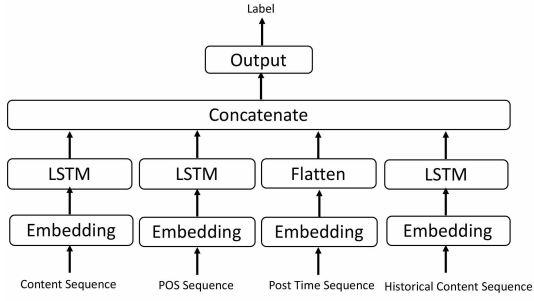}
\caption{Joint-LSTM for Text Classification}
\end{figure} 

Figure 2 shows an example design of Joint-LSTM model.
The sequence of part-of-speech tags and the post time of the tweet shown in the figure are the  {\em direct contextual features} which have direct relation with the target tweet. 
The POS tag sequence is generated from the word sequence, and it is fed into the model using embedding and LSTM layers.
Post time is a feature that is closely related to offline activities. 
We treat it as a sequence of size 1 to be able to use it flexibly in multiple models.
It turns out that there is little difference in terms of the overall performance between this approach as compared to other approaches such as directly feeding the time into a dense layer.
In addition, the J-LSTM model in Figure 2 also includes historical tweets.
They are modeled similar to the target tweet, and they share the same embedding layer with the target tweet.
Since the concatenation happens to the flat representation of the input sequences, J-LSTM suffers from the weakening of sequential information for the contextual and content features.

\subsubsection{Contextual-LSTM}
Ghosh {\em et al.} \cite{ghosh2016contextual} propose a Contextual LSTM (C-LSTM) model to handle contextual information.
They directly add the contextual feature to the decision function of each gate, as shown in the following equations.

\begin{align*}
i_t &= \sigma(W_{xi}x_t+W_{hi}h_{t-1}+W_{ci}C_{t-1}+b_i+\mathbf{W_{Ei}E})\\
f_t &= \sigma(W_{xf}x_t+W_{hf}h_{t-1}+W_{cf}C_{t-1}+b_f+\mathbf{W_{Ei}E})\\
c_t &= f_tC_{t-1}+i_ttanh(W_{xc}C_t+b_c+\mathbf{W_{Ei}E})\\
o_t &= \sigma(W_{xo}x_t+W_{ho}h_{t-1}+W_{co}c_t+b_o+\mathbf{W_{Ei}E})\\
h_t &= o_ttanh(c_t)
\end{align*}

where $i$, $f$ and $o$ are the input, forget,  and output gates, respectively, 
$x$ is the input, $c$ is the cell memory, $b$ is the bias, $h$ is the output, and $E$ is the contextual features.

\begin{figure}[htb]
\centering
\includegraphics[scale=0.4]{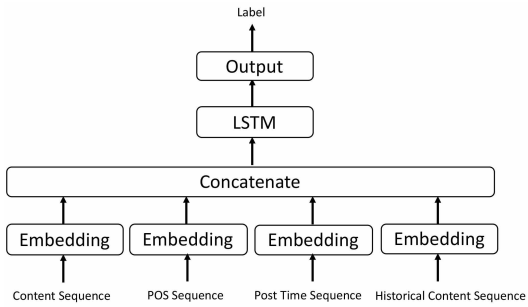}
\caption{Contextual-LSTM for Text Classification}
\end{figure}

The implementation of C-LSTM turns out to be quite simple.
It concatenates the embedded sequences of the contextual features with the embedded sequence of the content, and the concatenation is sent to an LSTM layer.
Figure 3 shows an example of C-LSTM model that takes POS sequence, post time sequence, and historical tweets as contextual features.
To properly form the concatenation with all the input embeddings, static features such as post time are duplicated and transferred into a sequence of the same value.
Using the same input and embedding settings as the J-LSTM model, the embeddings of the target tweet content and the contextual features are concatenated before sending to the LSTM layer.
Therefore, it is straightforward to see that C-LSTM requires the contextual features to have certain relationship with the content at every timestep. 

\subsubsection{Hierarchical-LSTM}
Existing Hierarchical LSTM (H-LSTM) models such as \cite{zhou2016hierarchical} are mainly used to model contents at different levels of details.
In addition, Huang {\em et al.} \cite{huang2016modeling} use the structure to incorporate social context such as retweets and replies.
In contrast, we utilize a similar H-LSTM structure, but include the historical tweets from the same author in a chronological order.

\begin{figure}[htb]
\centering
\includegraphics[scale=0.4]{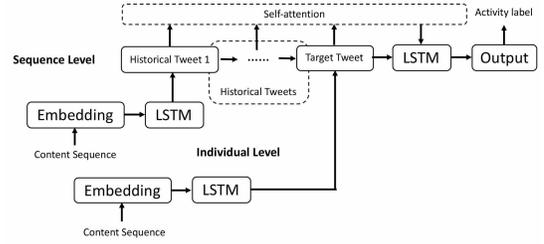}
\caption{Hierarchical-LSTM for Text Classification}
\end{figure}

Figure 4 shows the structure of the H-LSTM model.
Each LSTM segment on the individual level handles a single tweet sequence.
The input to the sequence level LSTM is a propagation of historical tweet representations where the first one being the oldest tweet and the last one being the target tweet.
Since the tweet representations in the sequence level are formed in a chronological order, the sequence can be modeled to learn the historical background towards the activity label of the target tweet.
To further help utilize the historical tweets, we also add a self-attention mechanism \cite{vaswani2017attention} to the LSTM on the sequence level.
All tweet contents share the same embeddings across the model.
The hierarchical structure strictly limits the type of features that can be used, therefore tests on other contextual features such as post time and POS tag sequence result in disappointing performances.

\section{Our Proposed Hybrid-LSTM Model}

\begin{figure*}[htb]
\centering
\includegraphics[scale=0.4]{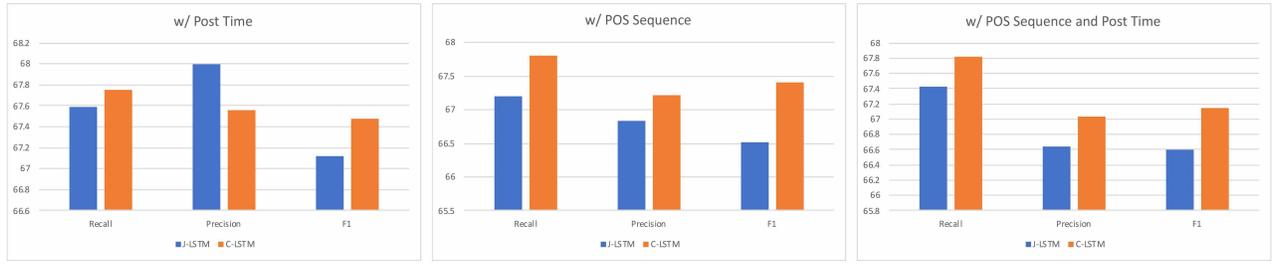}
\caption{Comparison between Different Models on Contextual Features}
\end{figure*}

\subsection{Including Historical Tweets} 
\begin{figure}[htb]
\centering
\includegraphics[scale=0.4]{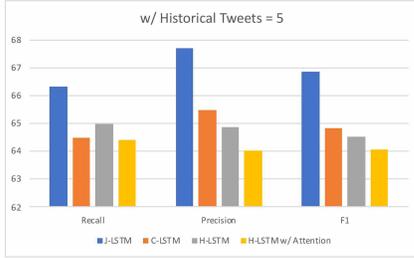}
\caption{Comparison between Different Models on Historical Features}
\end{figure}

In this section, we first analyze the three popular models described in the previous section  with respect 
to their ability to incorporate contextual features.
Based on the analysis, we  propose a hybrid LSTM model to better support  rich contextual learning.

We conduct a  comparison on a development dataset using J-LSTM, C-LSTM, and H-LSTM with features of POS tag sequence, post time, and historical tweets.
Details on the construction of the dataset will be covered in the experiment section.
These features are used to explore a more general conclusion for the capability of the contextual models.
The accuracies shown in Figure 5 and 6 are weighted averages across all labels to handle the imbalanced dataset.
In addition, Table II lists several sample tweets that will be used in the ensuing analysis.
\begin{table}[hbt]
\caption{Sample Tweets for Model Analysis}
\footnotesize
\centering
\begin{tabular}{c|l}
\hline
1&Nice day for a game. Less nice was Warren's first inning.\\
\hline
2&Biggest flag I've seen in person. Very cool. \#NeverForget \#911\\
\hline
3&We made it. \#BEmediaday\\
\hline
4&The wait is over! \#GreatBarrierReef \#Ashes \#GoldCoast\\
\hline
\end{tabular}
\end{table}

Figure 6 shows the use of three models in handling the most recent 5 historical tweets.
We test with different numbers of historical tweets, and found that the relative 
performances of different models are similar. 
Tweet 1 in Table II was posted while watching a baseball game, and the author only posts baseball related tweets.
It is surprising that H-LSTM has the worst performance as the structure is designed specifically for historical data. It cannot recognize the correct activity for Tweet 1 either.
Attention mechanism aims to handle historical information more appropriately, but it does not help generate any improvement.
The utilization of chronological order in including historical tweets may not be applicable to activity recognition on the target tweet.
In other words, the habit of posting tweets may not form a chronological dependency chain across historical tweets. 

C-LSTM incorporates historical information by a step-wise concatenation of the tweet sequences.
We believe that historical tweets have hidden information related to the target tweet, but such information is unlikely to be effectively captured in a word-to-word style.
Similar to C-LSTM, J-LSTM does not carry any order information.
The merging of the information for J-LSTM happens at the level of entire tweets.
So it relies on the sharing of the complete information among historical tweets.
Due to the fact that the historical tweets of Tweet 1 also contain a lot of baseball related words, J-LSTM and C-LSTM are able to recognize the correct activity of Tweet 1.
In addition, the historical tweets of Tweet 2 are very diverse in terms of the length, topic, and writing style.
Therefore, C-LSTM is not able to filter the noise while J-LSTM still works by combining the complete information.
Based on this analysis, we think that a simple combination of complete recent tweets could better support the classification of the target activity. 

\subsection{Including Direct Contextual Features}
Since H-LSTM is introduced to include historical tweets, we only apply J-LSTM and C-LSTM to the contextual features of POS tags and post time (Figure 5).
In general, C-LSTM performs better in handling both features.
Since it is designed to incorporate features at each step of the input sequence, it generates a larger improvement with step-wise features such as POS tags.
When dealing with static features like post time, it adds the same information to the gate decision for each input step of the content sequence.
On the other hand, J-LSTM incorporates this contextual information to the representation of the entire target tweet.

Tweet 3 is relatively short, but the post time of 6:19 a.m. would help to recognize the activity of traveling.
Meanwhile, after segmenting the hashtags in Tweet 4, knowing the tokens are proper nouns definitely help understand the author is traveling to Australia.
For both tweets, C-LSTM performs better by including the contextual information more accurately with the corresponding words.
Therefore, with deeper and more precise incorporation at each timestep, C-LSTM is more suitable in handling direct contextual features.

\subsection{Hybrid-LSTM}

The analyses above show that historical features are better handled by concatenation at the flat representation level and direct contextual features work better with step-wise concatenations.
In order to handle rich contextual learning that includes different types of contextual features, we propose a hybrid LSTM model (HD-LSTM) based on the analysis above.
HD-LSTM aims to cover a wide range of contextual features, and utilize different modeling layers for different contextual features.
With the capability of various layers in incorporating certain features, HD-LSTM is able to reach a better performance by handling the contextual features more appropriately. 

\begin{figure}[h]
\centering
\includegraphics[scale=0.4]{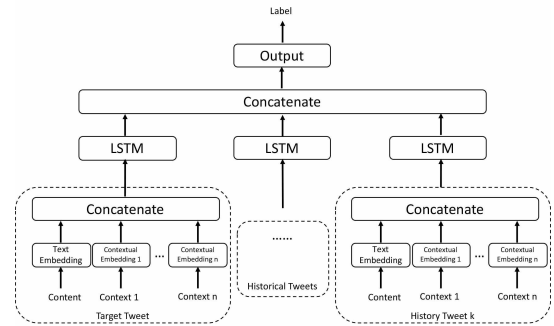}
\caption{Hybrid-LSTM for Text Classification}
\end{figure}

\begin{table*}[hbt]
\caption{Sample Tweets and Model Predictions}
\footnotesize
\centering
\begin{tabular}{c|l|c|c|c|c}
\hline
&\multicolumn{1}{|c|}{\textbf{Content}}&\textbf{True / Hybrid}&\textbf{LSTM}&\textbf{w/ Hist}&\textbf{w/ POS\&T}\\
\hline
1&Breakfast of champions&Traveling&Dining&Traveling&Dining\\
\hline
2&I guess the word has gotten out about E's ... so crowded today&Dining&Shopping&Dining&Dining\\
\hline
3&Last time I was here was pretty sad. \#BaptistHospital&Enhancement&Enhancement&Entertaining&Enhancement\\
\hline
\end{tabular}
\end{table*}

Figure 7 shows a sample design of HD-LSTM that takes text input, along with contextual features of historical information, POS tag sequence, and post time.
In particular, for each tweet component shown in the dashed box, the content sequence and the direct contextual features are combined with a concatenation of their embeddings.
In each dashed box, post time is used to mark the moment when the tweet was written, while POS tag sequence helps understand how each word was actually used in the tweet.
Then the enriched sequential representation is fed into a LSTM network and generates a flat vector representation for the tweet component.
At this step, each LSTM module learns the representation for the semantic, syntactic, and temporal information of the tweet. 
Next, the enriched flat representations for all tweets are concatenated to form a larger representation that contains the information from all inputs.
This concatenation further includes the historical information of the target tweet to improve the overall understanding of an enriched background.
Finally, the concatenated vector is fed into the output layer and generates the result label.

The features that belong to the same type across all tweet components share the same embedding.
In our case, all tweet content sequences, POS tag sequences, and post time share the same embeddings respectively.
To further boost the proposed hybrid model, we also add self-attention to all involved LSTM layers.

\subsection{Illustrative Examples}

Table III lists several examples from the development set to show the effect of including contextual features in recognizing activities, and the success of the proposed hybrid model.
We use LSTM to show the performance of using content only, use J-LSTM to apply historical tweets, use C-LSTM to include both POS tags and post time features, 
and use Hybrid LSTM to combine all these contextual features.

Tweet 1 shows a strong relation to breakfasts, however, the true situation is that the author took a photo of a sandwich while he is waiting at an airport.
It is reasonable that using the tweet content leads to a decision of ``dining'' activity, and it holds the same even if the post time is considered.
Meanwhile, the most recent two historical tweets from the author talked about leaving the hotel and arriving at the airport.
Thus, including the historical tweets become very useful in recognizing the correct ``traveling'' activity.
Tweet 2 describes a situation where the author is surrounded by many people.
With only this clue, it is possible that the author was shopping at a mall, having a dinner, or waiting at a train station.
Given the post time as 12:07 p.m. on a Sunday, it increases the possibility of having a meal, and the model is able to make the correct decision.
In fact, the true activity of the author is dining in a cafeteria, and ``E'' is the name of the place.
Since ``E'' is a very unusual name for a cafeteria, it becomes hard for content only model to utilize this information.
In addition, recent tweets from the author talk about having fun with friends, which also helps determine the correct activity. 
Tweet 3 has a strong indicator that the author is at a hospital, and the content only model can generate the correct output.
However, including only the historical tweets results in an incorrect result of ``entertaining''.
Several historical tweets are talking about drinking wine, which could mislead the historical model.
Those historical tweets are all posted at night, while the post time of the target tweet is early in the morning.
Considering this, the hybrid model is able to give the correct decision by distinguishing the different topics between the target tweet and the historical tweets.
\section{Experimental Results}

In this section, we describe our experiments that explore the performance of different LSTM based models, 
focusing on  comparing  their  abilities of incorporating contextual features towards a tweet classification task.
As we have stated throughout, the contextual features include POS tag sequence and post time of a tweet, as well as the most recent historical tweets from the same author.
Though author identity has been proved to be helpful in many tasks \cite{bakshy2011everyone,cha2010measuring}, 
 we do not include it  since it could potentially create a strong bias to the model and it is not general enough towards ordinary inference tasks.

\begin{table*}[hbt]
\caption{Location - Activity Label Mapping}
\footnotesize
\centering
\begin{tabular}{c|c|l}
\hline
\textbf{Activity}&\textbf{Tweet Count}&\textbf{Locations}\\
\hline\hline
Enhancement&3848&hospital, library, beauty salon, dentist, doctor, school, spa, university, physiotherapist\\
\hline
Traveling&12371&airport, bus station, train station, transit station, lodging, subway station\\
\hline
Dining&3934&bakery, liquor store, bar, restaurant, meal delivery, cafe\\
\hline
Entertaining&11457&amusement park, aquarium, movie theater, museum, zoo, park, casino, night club, art gallery\\
\hline
Shopping&4045&shopping mall, pharmacy, department\textbackslash book\textbackslash clothing\textbackslash pet\textbackslash convenience\textbackslash shoe\textbackslash electronics store\\
\hline
Sporting&10028&stadium\\
\hline
\end{tabular}
\end{table*}

\begin{table*}[ht]
\caption{Comparison of Model Performance}
\centering
\begin{tabular}{l|c|c|c|c|c|c|c|c|c|c|c}
\hline
&\multicolumn{4}{c V{3}}{\textbf{Content-only}}&\multicolumn{5}{c V{3}}{\textbf{J-LSTM}}&\multicolumn{2}{c}{\textbf{H-LSTM}}\\
\cline{2-12}
&\textbf{LSTM}&\textbf{BiLSTM}&\textbf{CNNLSTM}&\multicolumn{1}{cV{3}}{\textbf{LSTM+Att}}&\textbf{Time}&\textbf{POS}&\textbf{Direct}&\textbf{Hist=5}&\multicolumn{1}{cV{3}}{\textbf{All}}&\textbf{Hist=5}&\textbf{Hist=5+Att}\\
\hline
\textbf{Recall}&65.62&66.62&65.62&\multicolumn{1}{cV{3}}{66.99}&66.65&66.00&66.12&67.30&\multicolumn{1}{cV{3}}{66.91}&65.16&65.56\\
\textbf{Precision}&65.25&66.02&65.01&\multicolumn{1}{cV{3}}{66.66}&65.76&65.40&65.94&67.03&\multicolumn{1}{cV{3}}{67.88}&66.62&65.53\\
\textbf{F1}&64.96&65.71&65.06&\multicolumn{1}{cV{3}}{66.56}&65.98&65.54&65.98&67.04&\multicolumn{1}{cV{3}}{67.19}&65.69&65.44\\
\hline\hline
&\multicolumn{5}{c V{3}}{\textbf{C-LSTM}}&\multicolumn{4}{cV{3}}{\textbf{HD-LSTM w/ Hist=5}}&\multicolumn{2}{c}{ }\\
\cline{2-10}
&\textbf{Time}&\textbf{POS}&\textbf{Direct}&\textbf{Hist=5}&\multicolumn{1}{cV{3}}{\textbf{All}}&\textbf{Time}&\textbf{POS}&\textbf{Direct}&\multicolumn{1}{cV{3}}{\textbf{Direct+Att}}&\multicolumn{2}{c}{ }\\
\cline{1-10}
\textbf{Recall}&66.73&66.85&67.01&66.77&\multicolumn{1}{cV{3}}{66.80}&67.68&67.70&\textbf{68.70}&\multicolumn{1}{cV{3}}{\textbf{69.74}}&\multicolumn{2}{c}{ }\\
\textbf{Precision}&66.62&66.33&66.53&66.74&\multicolumn{1}{cV{3}}{67.61}&68.60&67.06&\textbf{68.13}&\multicolumn{1}{cV{3}}{\textbf{70.00}}&\multicolumn{2}{c}{ }\\
\textbf{F1}&66.29&66.30&66.61&66.33&\multicolumn{1}{cV{3}}{67.06}&68.03&67.22&\textbf{68.23}&\multicolumn{1}{cV{3}}{\textbf{69.84}}&\multicolumn{2}{c}{ }\\
\hline
\end{tabular}
\end{table*}

\subsection{Data preparation}

Manual labeling, though normally desirable for supervised learning, was problematic for labeling  tweets with activities for the following two reasons. 
First, humans are good at recognizing surface meaning, especially for the case that no background and external information are required.
Thus, manual labeling suffers from the same problem showed by the examples described in the first section.
The activities that cannot be inferred from the content itself  are unlikely to be correctly labeled by humans.
Second, labeled dataset of sufficient size was highly desirable, as the size of the training data is highly related to the quality of the model. 
Although there are certain ways to crowdsource the labeling process, obtaining sufficient labeled tweets with consistent quality seemed infeasible. 
Therefore, we label the activities based on the reported locations.

We started the data collection from defining a list of place categories that are strongly related to certain activities.
Then we used Google Maps API to collect specific places for each category with detailed coordinates.
Finally, we used Twitter API to collect tweets that are posted with a reported location,  which is also in  a range of 10 meters from the coordinates of a specific place.
We removed duplicates and only included the tweets that have reported location type as  {\em Point of Interest}  (POI).
POI indicates that an activity can be conducted at this location \cite{lian2011collaborative}.
To further clean the data, we removed tweets that contain less than 3 tokens or more than 70\% of the tokens are mentioned usernames.
Hashtags are useful elements in tweets, and sometimes they can be strong indicators for locations or activities.
However, such use of hashtags may also lead to overfit the model, and the uniqueness of creating hashtags makes it less useful towards unseen ones.
To prevent this problem while preserve the meaning, we removed the hashtag signs and segmented the hashtag content so that the hashtags are separated into ordinary words.

Table IV shows the relationship between the predefined place categories and activities.
As mentioned, additional rules are used to improve the labeling quality, such as tweets that have noun keyword ``ceremony'' at location ``stadium'' should be labeled as ``enhancement''. 

Although the data collection process is initialized with the same amount of requests for each activity type, it results in an imbalanced dataset.
In our test, down-sampling or over-sampling the dataset does not show any considerable difference in the overall performance.
Therefore, training data are processed with different weights with respect to different classes,
and the metrics are calculated as the weighted average across classes (one consequence is that F1-score may not fall in between precision and recall values).
The training, development, and test sets are randomly divided with ratios of 0.6, 0.2 and 0.2.

\subsection{Experiment Settings}

To show the improvement of using contextual features, we also experiment with other content-only LSTM based models, i.e., BiLSTM \cite{schuster1997bidirectional}, CNNLSTM \cite{zhou2015c}, and LSTM with self-attentions (LSTM+Att).
Unlike certain previous tasks \cite{vosoughitweet2vec,dhingratweet2vec}, using word-level model results in a better performance than character-level model in our task.
We apply the idea of transfer learning to initialize tweet content embeddings using GloVe \cite{pennington2014glove} before training.
This creates a more domain-specific word embedding compared with using fixed pre-trained embeddings, and it also generates better performance compared with randomly initialized embeddings.
Additionally, POS embeddings are randomly initialized.
Post time is represented as day of the week and time of the day, and we set four time periods for a day.
Tweet content embeddings have 200 dimensions, while POS tags, time, and day are all mapped to embeddings of 20 dimensions.

Testing with different numbers of historical tweets, we found that including 5 most recent tweets as the contextual feature yields the optimal performance for most models.
It should be pointed out that H-LSTM is much more sensitive to the number of historical tweets compared with other models. 
POS tags are generated using a tweet-specific tagger \cite{owoputi2013improved}, and the models are mainly built using Keras \cite{Charles2013} \footnote{Source code is available at https://goo.gl/o9dsBh}.
We use 200 nodes for all the LSTM networks in the experiment with a dropout rate of 0.2, categorical cross-entropy as the loss function, 
apply Adam optimization for training, and set a mini-batch of size 100.
Softmax function is used in all output layers, and all models are tuned with different epochs for optimal performance.

\subsection{Model Performance}

\begin{figure*}[ht]
\centering
\includegraphics[scale=0.73]{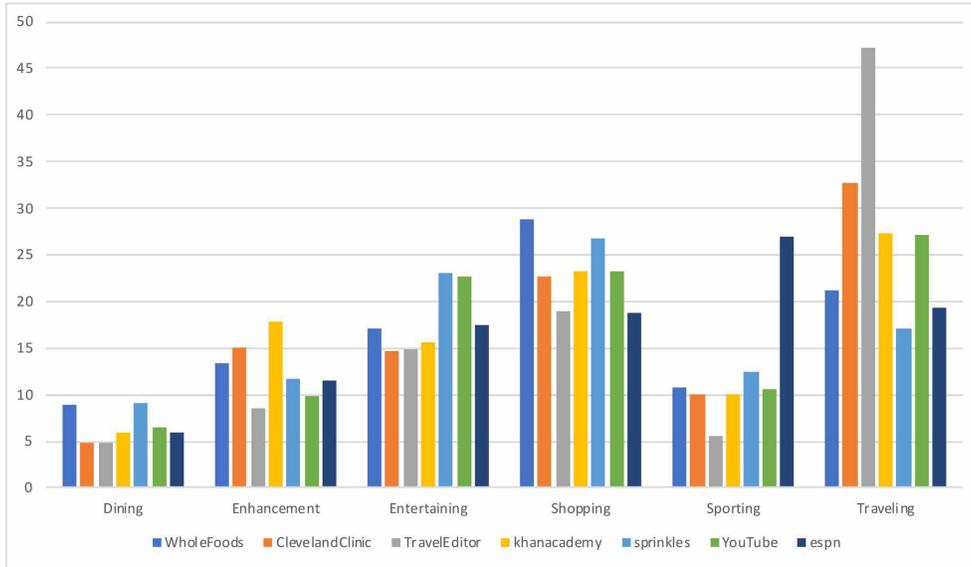}
\caption{Summary of the Activity Distributions for Followers of Certain Accounts}
\end{figure*}

Table V lists the performance of different models.
For contextual features, ``Direct'' refers the use of POS tag sequence and post time features in addition to target tweet content, 
while ``All'' denotes the use of POS sequence and post time with the content of both target tweet and 5 most recent historical tweets.

Models that only use the target tweet content results in generating only limited improvement over the original LSTM.
In contrast, the use of contextual features boosts the performance.
The post time is more useful than the POS tag sequence, and the benefit of including historical tweets varies with the method of incorporation.

LSTM uses only the content of tweets and reaches a reasonable performance for the task given it has 6 labels.
Bi-LSTM adds the ability to understand the content in another order and helps improve the outcome.
Meanwhile, adding the convolutional layer does not provide much improvement. 
CNN is used to extract information similar to an n-gram model, and the informal use of words in tweets reduces the capability of such information.
As expected, adding attention mechanism considerably helps the performance.

J-LSTM works better to include historical tweets and C-LSTM performs better with direct contextual features, while H-LSTM does not do well to include historical tweets.
Since C-LSTM incorporates the contextual features into every token of the input sequence, C-LSTM shows to benefit from adding more direct contextual features.
All three contextual models are able to benefit from including historical tweets.
It is surprising that C-LSTM generates a certain level of improvement with historical tweets.
C-LSTM includes the tokens from historical tweets with the tokens from the target tweet at each time step, 
and it is not intuitively correct that words from different tweets have direct relationships.
We think that some hidden attributes across tweets from the same author bring the improvement, such as the use of certain words while the author is engaged in a particular activity.

Combining the power of both J-LSTM and C-LSTM, the hybrid model outperforms both content-only models and models that use a fixed method to incorporate contextual features.
When including all features, the large improvement of HD-LSTM over J-LSTM and C-LSTM shows the effectiveness of the hybrid model. 
The reported performance improvements further strengthen the analysis that was used to build the proposed model: 
historical tweets can be better handled by concatenating the complete information of tweets, and the step-wise concatenation of feature representations works better to include direct contextual features.
It is also obvious that HD-LSTM benefits from simply including more contextual features. 
In contrast, using a single method to incorporate more contextual features does not consistently improve the performance.
Finally, HD-LSTM also benefits from adding self-attention mechanism to LSTM layers. 


\section{Demonstrating Use of the Approach: A Case Study} 

In this section, we exhibit a real case where the activity recognition is utilized on a large volume of tweets.
The results validate the effectiveness of the activity recognition model.

We find 7 popular accounts that all have a large number of followers but are distinct with 
their fields of focus.
For each account, we collect 10,000 followers randomly, and for each follower, we collect the most recent 200 tweets.
For each tweet, we apply he hybrid model with POS sequence, post time and historical tweet features to generate a probability distribution over activities.
Then we generate a distribution of activities for each follower by combining the distributions of the tweets posted by that follower. 
Thus, we are able to accumulate the distribution for each follower to generate a probability distribution of the activity labels over the collection of followers for each popular account.
This activity distribution is used to represent the follower activity profile for this popular account.
$p_{f,t,i}$ is the probability for the $i_{th}$ activity label given a single tweet $t$ from follower $f$, and the probability $P_i$ for the $i_{th}$ activity for the collection of followers for an account would be:

\begin{align*}
P_i &=\frac{1}{Z_0}\sum_{f\in F}\frac{1}{Z_1}\sum_{t\in T}p_{f,t,i}\\
\end{align*}

Since there are duplications and invalid tweets involved in the dataset, the number of tweets for each follower used for the model may not be the same.
Therefore, we have a normalization factor $Z_1$ to normalize for each follower, and another factor $Z_0$ to normalize for each popular account.
In addition, $F$ is the set of followers for the account, and $T$ is the collection of tweets for a particular follower.

We train the model using the full dataset from the experiment, and Figure 8 shows the analysis result for these popular accounts.
To make the graph more understandable, we present the probability for each activity label over popular accounts, so the probabilities for each activity label do not sum up to 1.
The imbalanced dataset used to train the model creates certain trends in different activity labels, but the comparison within each activity label can still be useful to draw some conclusions.

It is straightforward to see that {\em espn} has a high probability for ``Sporting'' and {\em TravelEditor} holds the peak in ``Traveling''.
{\em khanacademy} and {\em ClevelandClinic} represent educational and medical needs, and it leads to a obvious result of the highest probabilities in ``Enhancement''.
It is interesting that {\em ClevelandClinic} has the second highest amount of attention of its followers for travel.
The need of expanding medical services from the team and the need of heading to medical facilities from the patients could cause such increasing attention in ``Traveling''.
{\em WholeFoods} and {\em sprinkles}, as a food market chain and famous cupcake bakery, have the highest involvement of both ``dining'' and ``shopping'' for their followers.
It shows that the followers of {\em WholeFoods} also care about personal enhancement other than foods.
{\em YouTube} has a high involvement of ``Entertaining'' for its followers, while the peak of {\em sprinkles} indicates that the interest in cupcakes could lead to the interest in entertainment.

These observations and conclusions are another types of validation that shows the usefulness and effectiveness of the activity recognition model.
\section{Conclusions and Future Work}

We have presented a methodology for  including contextual features to improve the performance of content-based LSTM models,
with an application of recognizing  offline activities of a user when posting the tweets.
Our contributions include a location-based method to label tweets with offline activities, as well as an analysis and exploration of the different ways of including direct and historical contextual features with LSTM and each technique's effectiveness.
Then we propose a hybrid LSTM model that combines and takes advantage of the various methods to include contextual features.
Our experiments show that including contextual information can easily outperform the content only models, 
and our hybrid model is able to incorporate the contextual features more effectively than existing methods.
The amount of improvement shows the importance of choosing the right method for including certain types of contextual features.
Finally, we validate our activity recognition model by using it to derive the activity analysis of the followers for several popular Twitter accounts. 

We intend to identify more contextual features and explore their abilities using additional models.
Our current labeling process highly relies on the reported location of each tweet, 
thus figuring out better ways to improve location accuracy could potentially increase the usefulness of our work. 
During our experiments, we found that some images attached to the tweet may be useful in identifying activities.
While it is currently not common to include images as contextual features for text data,  we believe this could be a promising direction of research.

\bibliographystyle{IEEEtran}
\bibliography{reference}

\end{document}